\documentclass[letterpaper,twocolumn,prl,aps,superscriptaddress,amsmath,amssymb,floatfix]{revtex4-1}
\usepackage{mathptmx}
\usepackage[latin9]{inputenc}
\usepackage{color}
\usepackage{amsmath}
\usepackage{amssymb}
\usepackage{graphicx}
\usepackage{esint}
\usepackage[unicode=true,
 bookmarks=true,bookmarksnumbered=false,bookmarksopen=false,
 breaklinks=false,pdfborder={0 0 1},backref=false,colorlinks=true]
 {hyperref}
\hypersetup{
 linkcolor=magenta,urlcolor=blue,citecolor=blue,pdfstartview={FitH},hyperfootnotes=false}

\makeatletter

\usepackage{textcomp}
\usepackage{epstopdf}

\usepackage{amsfonts}

\pdfpageheight\paperheight
\pdfpagewidth\paperwidth

\@ifundefined{textcolor}{}{%
 \definecolor{BLACK}{gray}{0}
 \definecolor{WHITE}{gray}{1}
 \definecolor{RED}{rgb}{1,0,0}
 \definecolor{GREEN}{rgb}{0,1,0}
 \definecolor{BLUE}{rgb}{0,0,1}
 \definecolor{CYAN}{cmyk}{1,0,0,0}
 \definecolor{MAGENTA}{cmyk}{0,1,0,0}
 \definecolor{YELLOW}{cmyk}{0,0,1,0}
}

\usepackage{xcolor}\usepackage{soul}
\definecolor{blue}{rgb}{0,0,1}
\definecolor{red}{rgb}{1,0,0}
\definecolor{green}{rgb}{0,1,0}

\makeatother

\begin{document}
\title{Break the efficiency limitations of dissipative Kerr soliton using nonlinear couplers}
\author{Ming Li}
\affiliation{Key Laboratory of Quantum Information, CAS, University of Science
	and Technology of China, Hefei, Anhui 230026, China}
\affiliation{CAS Center For Excellence in Quantum Information and Quantum Physics,
	University of Science and Technology of China, Hefei, Anhui 230026,
	China.}
\author{Xiao-Xiao Xue}
\affiliation{Department of Electronic Engineering, Beijing National Research Center
	for Information Science and Technology, Tsinghua University, Beijing,
	China}

\author{Yan-Lei Zhang}
\affiliation{Key Laboratory of Quantum Information, CAS, University of Science
	and Technology of China, Hefei, Anhui 230026, China}
\affiliation{CAS Center For Excellence in Quantum Information and Quantum Physics,
	University of Science and Technology of China, Hefei, Anhui 230026,
	China.}
\author{Xin-Biao Xu}
\affiliation{Key Laboratory of Quantum Information, CAS, University of Science
	and Technology of China, Hefei, Anhui 230026, China}
\affiliation{CAS Center For Excellence in Quantum Information and Quantum Physics,
	University of Science and Technology of China, Hefei, Anhui 230026,
	China.}
\author{Chun-Hua Dong}
\email{chunhua@ustc.edu.cn}

\affiliation{Key Laboratory of Quantum Information, CAS, University of Science
	and Technology of China, Hefei, Anhui 230026, China}
\affiliation{CAS Center For Excellence in Quantum Information and Quantum Physics,
	University of Science and Technology of China, Hefei, Anhui 230026,
	China.}

\author{Guang-Can Guo}
\affiliation{Key Laboratory of Quantum Information, CAS, University of Science
	and Technology of China, Hefei, Anhui 230026, China}
\affiliation{CAS Center For Excellence in Quantum Information and Quantum Physics,
	University of Science and Technology of China, Hefei, Anhui 230026,
	China.}
\author{Chang-Ling Zou}
\email{clzou321@ustc.edu.cn}

\affiliation{Key Laboratory of Quantum Information, CAS, University of Science
	and Technology of China, Hefei, Anhui 230026, China}
\affiliation{CAS Center For Excellence in Quantum Information and Quantum Physics,
	University of Science and Technology of China, Hefei, Anhui 230026,
	China.}
\date{\today}

\begin{abstract}
Dissipative Kerr soliton (DKS) offers a compact solution of coherent comb sources and holds huge potential for applications, but has long been suffering from poor power conversion efficiency when driving by a continuous-wave laser. Here, a general approach to resolving this challenge is provided. By deriving the critical coupling condition of a multimode nonlinear optics system in a generalized theoretical framework, two efficiency limitations of the conventional pump method of DKS are revealed: the effective coupling rate is too small and is also power-dependent. Nonlinear couplers are proposed to sustain the DKS indirectly through nonlinear energy conversion processes, realizing a power-adaptive effective coupling rate to the DKS and matching the total dissipation rate of the system, which promises near-unity power conversion efficiencies. For instance, a conversion efficiency exceeding $90\:\%$ is predicted for aluminum nitride microrings with a nonlinear coupler utilizing second-harmonic generation. The nonlinear coupler approach for high-efficiency generation of DKS is experimentally feasible as its mechanism applies to various nonlinear processes, including Raman and Brillouin scattering, and thus paves the way of micro-solitons towards practical applications.
\end{abstract}
\maketitle

\section{Introduction}

During the last decades, soliton frequency comb
based on Kerr medium has attracted tremendous research interests in
the communities across fiber optics and integrated photonics\,\cite{fortier201920,kippenberg2018dissipative}. Phase-locked comb states including single-soliton state\,\cite{PhysRevA.89.063814} and soliton crystals\,\cite{Cole2017,karpov2019dynamics,Lu2021a} have been experimentally
observed in microcavities made by various $\chi^{(3)}$ materials,
such as silicon nitride (SiN)\,\cite{moss2013new,PhysRevLett.114.053901}, magnesium fluoride  ($\mathrm{MgF_{2}}$)\,\cite{herr2014temporal}, aluminum nitride (AlN)\,\cite{gong2018high}, silica\,\cite{yi2015soliton}, and lithium nitride (LN)\,\cite{gong2020photonic,gong2020near,he2020perfect}. They can be excited by pumping a selected resonance in a dispersion-engineered microcavity with only continuous-wave (CW) lasers under proper dynamical controlling\,\cite{guo2017universal,xue2015mode}. The DKS providing the wide-spread and phase-locked optical comb allows a wide range of applications in light
sources\,\cite{pasquazi2018micro,griffith2015silicon}, optical spectroscopy\,\cite{foltynowicz2011optical,picque2019frequency},
classical and quantum communications\,\cite{marin2017microresonator,wang2020quantum},
astronomy\,\cite{obrzud2019microphotonic}, ranging\,\cite{ghelfi2014fully}, microwave-to-optical link\,\cite{liu2020photonic,spencer2018optical}, optical clock\,\cite{Papp:14}, and
the emerging research field of machine learning\,\cite{Feldmann2021}.

However, only a very small portion of CW drive power can be converted
to the soliton. Introducing the power conversion efficiency $\eta$,
as the ratio of the output soliton comb power to the input power of
CW laser, the achievable $\eta$ is usually limited to several percent. What's worse, $\eta$ scales inversely with the pump power
and the number of comb lines, which is extremely low for soliton states
with large spectral width\,\cite{bao2014nonlinear}. The low conversion efficiency imposes a barrier for practical applications
of DKS comb and demands high-power pump lasers.
To achieve a high $\eta$, the microcavities working in the over-coupling regime are used
for increased energy coupling and extraction efficiencies\,\cite{jang2020universal},
which can be realized by designing the waveguide coupler to the cavity or using an additional cavity as ancillary\,\cite{xue2019super}.
Besides, a moderate $\eta \sim 0.17$ was demonstrated in Pockels
soliton comb using $\chi^{(2)}$ nonlinearity\,\cite{Bruch2020}. Beyond the CW pump, the pulse-pump approach is also developed to enhance the field overlap
between the input and intracavity fields for higher $\eta$\,\cite{obrzud2017temporal,malinowski2017optical,anderson2018achieving}. Despite the exciting results of these ingenious designs, the physical mechanism that limits the soliton conversion efficiency remains unclear, so is the
optimal condition for high conversion efficiency. Moreover, the fundamental
problem that the conversion efficiency decreases with the pump power
is not resolved either.

In this work, we develop a general theoretical framework of multi-channel input-output relation for multimode nonlinear optics system and reveal the underlying physics that limits the power conversion efficiency of DKS. The nonlinear coupler is proposed to break the limitations by engineering the coupling channels to the soliton state. Two types of nonlinear couplers are investigated. The nonlinear coupler for collective enhancement could be realized via optical parametric oscillation (OPO),
%provides a linear enhancement factor on the excitation rate and also 
which enables the simultaneous energy transfer from a monochromatic input field to soliton state via multiple comb lines of distinct frequencies. The self-adaptive nonlinear coupler based on second-harmonic generation (SHG) allows the pump-power-independent
high conversion efficiency. Our analytical derivation and analysis agree excellently with numerical simulations and show the experimental feasibility of our schemes. Our work opens new opportunities in multimode nonlinear systems for high-efficiency frequency conversion processes.

\section{Generalized critical coupling condition}

\begin{figure*}[t]
	\begin{centering}
		\includegraphics[width=0.9\textwidth]{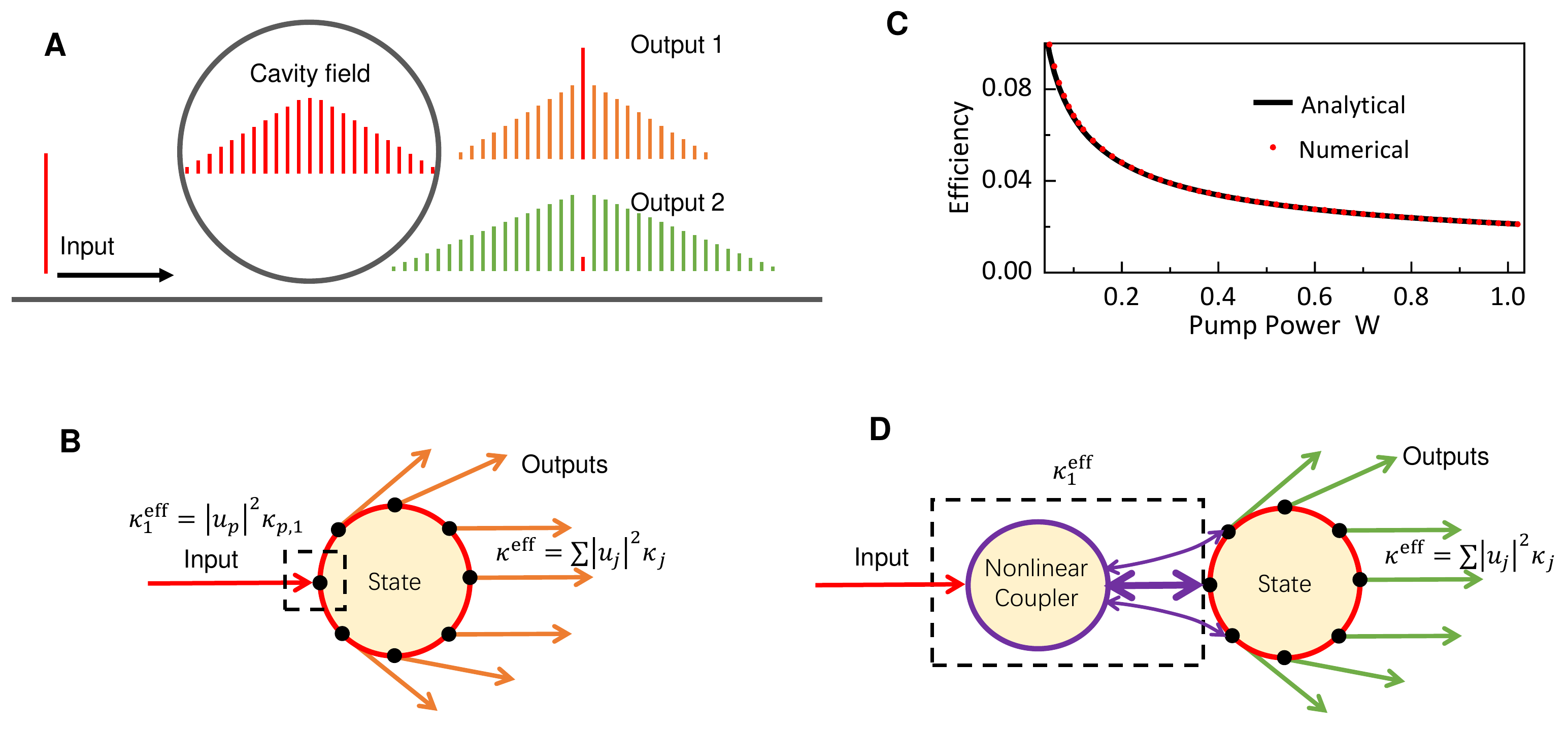}
		\par\end{centering}
	\caption{\textbf{Comb formation and extraction in a microring cavity pumped by a
			CW laser.} (\textbf{A}) For cavity modes uniformly coupled to the waveguide,
		the transmitted field (Output $1$) has a distinguished peak due to the residue of the
		CW pump laser, indicating a poor conversion efficiency. For an ideal case, the transmitted
		field (output $2$) shows a spectrum with most of the pump
		laser being absorbed by the cavity and converted to other frequencies.
		The intracavity field shows a typical $\mathrm{sech^{2}}$-shape envelope for the soliton state.
		(\textbf{B}) Schematic of the mechanism that limits the soliton conversion efficiency. The soliton state (red circle), as a collective of nodes (black dots) with each node representing an optical mode, couples with the CW input field through a single channel and dissipates through every node. 
		(\textbf{C}) Relationship between the conversion efficiency of a single-soliton
		state against the pump power for a typical AlN microcavity, with all modes are uniformly and critically coupled to a waveguide coupler. (\textbf{D}) Schematic of the nonlinear coupler, which induces the indirect coupling between the soliton state and input field, with the effective coupling rate of a selected channel bening enhanced and also multiple coupling channels are opened.}
	
	\label{Fig1}
\end{figure*}

\noindent Figure\,\ref{Fig1}(A) illustrates the formation of frequency combs inside an optical Kerr cavity, which is described by the Hamiltonian\ \cite{guo2018efficient}
\begin{eqnarray}
	H_{\mathrm{Kerr}} & = & \sum\Delta_{j}a_{j}^{\dagger}a_{j}+g_{3}\sum_{ijkl}a_{j}^{\dagger}a_{i}^{\dagger}a_{k}a_{l}\delta_{ijkl}.\label{eq:TotalH-1}
\end{eqnarray}
Here $a_{j}^{\dagger}$ ($a_{j}$) is the creation (annihilation)
operator of mode $j$ with $j\in\{-N,...0,...N\}$, $g_{3}$ is the
single-photon coupling strength of four-wave mixing (FWM) due to the
Kerr nonlinearity, $\Delta_{j}$ is the relative mode frequency detuning with respect to an equally-spaced frequency comb, i.e. in the rotating frame of $\sum\left(\omega_{p}+jD_{1}\right)a_{j}^{\dagger}a_{j}$, $D_{1}$ and $\omega_{p}$ are the cavity free-spectral range and the frequency of the pump laser, respectively. $\delta_{ijkl}=1$ iff $i+j=k+l$ or else $\delta_{ijkl}=0$.

Instead of studying the dynamics of each mode, we treat the multimode
comb state $\overrightarrow{\boldsymbol{\alpha}}=\{\alpha_{j}\}$
as a collective state, with $\alpha_{j}$ denoting the field amplitude
of $j$-th mode and $\left|\alpha_{j}^{2}\right|$ corresponding to
the intracavity photon number. Then we investigate the effective coupling between the collective state $\overrightarrow{\boldsymbol{\alpha}}$ and the
input laser, and also consider its dissipation to the environment. As schematically illustrated by Fig.\,\ref{Fig1}(B), the comb state (red circle) is represented by a collection of modes (black dots), and it could be excited through selected modes by input lasers of corresponding frequencies and decays simultaneously from all modes. Due to the photon-number preserving nature of FWM, the interaction Hamiltonian [Eq.~(\ref{eq:TotalH-1})] only re-distribute photons among different modes while have no net contribution to the total photon number $N=\sum_{n}\langle a_{n}^{\dagger}a_{n}\rangle$ inside the cavity. Therefore, the photon flux dynamics (PFD) of the system follows
{[}see the Supplementary Materials for details{]}
\begin{eqnarray}
	\frac{dN}{dt} & = & -\sum_{j}2\kappa_{j}|\alpha_{j}|^{2}+S_{\mathrm{in}},\label{eq:EnergyDyn}
\end{eqnarray}
where $S_{\mathrm{in}}=2\mathrm{Re}[\overrightarrow{\varepsilon}\cdot\overrightarrow{\boldsymbol{\alpha}}^{*}]$ is the source term as a inner product of the input laser amplitudes and the corresponding mode amplitudes, and denotes the input photon flux of the system.
Here, $\overrightarrow{\boldsymbol{\varepsilon}}=\{\sqrt{2\kappa_{j,1}F_{j}}\}$ is the pump field vector with $F_{j}=P_{\mathrm{in},j}/\hbar\omega_{j}$,
$P_{\mathrm{in},j}$ and $\omega_{j}$ are the power and frequency of the input field on the $j$-th mode, respectively. Then, for a normalized state $\overrightarrow{\boldsymbol{\alpha}}=\sqrt{N}\overrightarrow{\boldsymbol{u}}$
with $\left|\overrightarrow{\boldsymbol{u}}\right|^{2}=1$, we have
\begin{equation}
	\frac{d}{dt}N=-2\kappa^{\mathrm{eff}}N+2\sqrt{2\kappa_{1}^{\mathrm{eff}}F_{\mathrm{tot}}}\sqrt{N},\label{eq:dynamics_eff}
\end{equation}
with an effective total dissipation rate $\kappa^{\mathrm{eff}}=\sum_{j}\kappa_{j}\left|u_{j}^{2}\right|$ and an effective external coupling rate 
\begin{eqnarray}
	\kappa_{1}^{\mathrm{eff}} & = & \left(\mathrm{Re}\sum_{j}\sqrt{\kappa_{j,1}F_{j}/F_{\mathrm{tot}}}u_{j}^{*}\right)^{2}\label{eq:k1eff}
\end{eqnarray}
from the collective driving with total input $F_{\mathrm{tot}}=\sum_{j}\left|F_{j}\right|$.
To achieve the highest conversion efficiency, it is usually required each $F_{j}$ is completely absorbed by the corresponding $j$-th mode. 

By applying the well-known single-mode input-output formalism $a_{j}^{out}=a_{j}^{in}-\sqrt{2\kappa_{j,1}}\alpha_{j}$
on each pumped mode$\:$\cite{walls2007quantum}, we obtain the
condition for achieving the optimal conversion efficiency of the collective state $\overrightarrow{\boldsymbol{\alpha}}$
\begin{eqnarray}
	2\kappa_{1}^{\mathrm{eff}} & = & \kappa^{\mathrm{eff}}\label{eq:ccm}
\end{eqnarray}
without solving the detailed dynamics of individual modes. Interestingly, the generalized
critical coupling condition of the multimode field in Eq.\,(\ref{eq:ccm})
is reminiscent of the critical coupling condition of a single mode and it
reduces exactly to the single-mode
case when the nonlinear interaction is absent.

\section{Soliton conversion efficiency}

Although the general model
reveals the underlying physics for the high-efficiency conversion of the collective state, the universal optimal coupling condition [Eq.~(\ref{eq:ccm})] requires the precise solution of $\overrightarrow{\boldsymbol{\alpha}}$, which is demanding since the solution varies with the input condition. Luckily, due to the balance between the second-order dispersion $D_{2}$ and Kerr nonlinearity $g_{3}$, the DKS states have the amplitude of the pump mode being bounded below $\alpha_{p}=\sqrt{\frac{D_{2}}{4g_{3}}}$ under certain approximation~\cite{herr2014temporal}.
Therefore, with only $\alpha_{p}$, we can re-derive the PFD {[}Eq.\,(\ref{eq:EnergyDyn}){]} and solve the steady state by
\begin{equation}
	0=-2\kappa_{j}\sum_{j\neq p}|\alpha_{j}|^{2}-2\kappa_{p}|\alpha_{p}|^{2}+2\mathrm{Re}[\varepsilon_{p}\alpha_{p}^{*}].
\end{equation}
So, for a given $\alpha_{p}$, the DKS is special that the output photon flux of the comb 
\begin{equation}
	S_{\mathrm{out}}=2\kappa_{a,1}\sum_{j\neq p}|\alpha_{j}|^{2}
\end{equation}
could be derived without knowing the exact soliton energy distribution $\overrightarrow{\boldsymbol{\alpha}}$. The analytical expression of the achievable conversion efficiency can be derived as (see the Supplementary Materials for details)
\begin{eqnarray}
	\eta & = & \frac{\kappa_{a,1}}{\kappa_{a}}\left[\frac{\kappa_{p,1}}{\kappa_{p}}-\left(\sqrt{\frac{\kappa_{p,1}}{\kappa_{p}}}-\sqrt{\frac{2\kappa_{p}\alpha^{2}}{P_{\mathrm{in}}/\hbar\omega_{p}}}\right)^{2}\right].\label{eq:efflinear}
\end{eqnarray}

According to Eq.~(\ref{eq:efflinear}), the conversion efficiency $\eta$ can be factorized to the product of the comb extraction efficiency $\eta_{ex}=\frac{\kappa_{a,1}}{\kappa_{a}}$
and the comb excitation efficiency (terms inside the bracket). The extraction efficiency $\eta_{ex}$ builds a universal bound of $\eta$ for any scheme and can only be increased at the expense of high comb generation threshold which scales with $\kappa_{a}^{2}$. Under the critical coupling condition of Eq.\,($\ref{eq:ccm}$), the
optimal conversion efficiency
\begin{eqnarray}
	\eta_{max} & = & \frac{\kappa_{a,1}}{\kappa_{a}}\frac{\kappa_{p,1}}{\kappa_{p}},\label{eq:maxefficiency-1}
\end{eqnarray}
can be achieved when the excitation efficiency reaches
its maximum $\frac{\kappa_{p,1}}{\kappa_{p}}$ for $\sqrt{2\kappa_{p,1}}\alpha=\sqrt{P_{\mathrm{in}}/\hbar\omega}$. At the same condition, the pump laser should be completely absorbed due to the input-output formalism for the pump mode, which indicates all pump laser power is converted to soliton and validates the self-consistence of our analysis.

As shown in Fig.\,\ref{Fig1}(C), the CW pump to single-soliton state conversion is studied in a typical microring cavity made by AlN~\cite{Bruch2020}. In the model, $D_2/2\pi=250\,\mathrm{MHz}$, $g_3/2\pi=2\,\mathrm{Hz}$, and all modes are critically coupled to the waveguide and have the same dissipation rate. Our analytical prediction agrees excellently with the numerical results for various pump powers, which confirms our generalized theory of the collective state. However, both approaches predict a very low conversion efficiency around $5\%$, and the efficiency even decreases with the pump power. 

Such limitations of the efficiency, i.e. the low efficiency and power-dependence, could be attributed to the strong deviation from the generalized critical coupling condition [Eq.~(\ref{eq:ccm})]: For a broad soliton comb with a single continuous input to the $p$-th mode, we have $\kappa_{1}^{\mathrm{eff}}=\kappa_{p,1}\left|u_{p}^{2}\right|\ll\kappa_{p,1}$.
Assuming that the comb modes have uniform loss rates $\kappa_{j}=\kappa$,
the decay rate of the soliton state $\kappa^{\mathrm{eff}}=\kappa\gg\kappa_{1}^{\mathrm{eff}}$.
As a result, the system works in the deep under-coupled regime with
most of the input light is directly transmitted through the cavity,
leading to very low conversion efficiency. What's worse, as the
comb becomes broad with increased pump power, the normalized field
of the pump mode $\left|u_{p}^{2}\right|$ decreases, so is the conversion
efficiency.

To break the efficiency limitation, we can design the external
coupling rate $\kappa_{p,1}$ to match the critical coupling condition in Eq.\,($\:$\ref{eq:ccm}). Submitting $\kappa_{1}^{\mathrm{eff}}=\kappa_{p,1}\left|u_{p}^{2}\right|$
into Eq.\,($\:$\ref{eq:ccm}), we obtain
\begin{eqnarray}
	\kappa_{p,1} & = & \kappa_{p,0}+\sum_{j\neq p}\left|\frac{\alpha_{j}}{\alpha_{p}}\right|^{2}\kappa_{j}.\label{eq:ccs}
\end{eqnarray}
It can be inferred from Eq.\,($\:$\ref{eq:ccs}), the pump mode experiences
intrinsic mode dissipation from itself by $\kappa_{p,0}$ and also
additional loss channels via the $j$-th mode by a weighted rate $\left|\frac{\alpha_{j}}{\alpha_{p}}\right|^{2}\kappa_{j}$
{[}Fig.\,\ref{Fig1}(B){]}.
This condition demands the external coupling rate on the pump mode
to be much larger than other comb modes. Such a strong mode-selective
coupler could be realized with a narrow-band coupler by introducing
a variety of dispersive photonic structures, such as gratings,
Mach-Zehnder interferometers~\cite{gong2020photonic} and resonators.
For example, an ancillary cavity was used to pump the soliton cavity
and demonstrated a high conversion efficiency in Ref.~\cite{xue2019super}.
Since the free spectral range of the two cavities are different, the
ancillary cavity can only mediate the coupling between one cavity
mode and the waveguide, thereby realizing the independent control
on the external coupling rate of the pump mode. It should be noted that, according to our analytical derivation (see
the Supplementary Materials), this scheme also shares the same conversion efficiency $\eta_{max}$ in Eq.\,(\ref{eq:maxefficiency-1})
for a pump-power dependent linear coupling rate $g_{l}=\frac{\varepsilon}{2\alpha}$
between the two cavities, thus it could be included in our current
unified framework.

From the generalized critical coupling condition of the collective state [Eq.~(\ref{eq:ccm})], a high conversion efficiency is only possible by significantly enhancing the $\kappa_{1}^{\mathrm{eff}}$. In addition to designing a mode-selective coupler, a multiple channel input field allows the optimal $\kappa_{1}^{\mathrm{eff}}$ in Eq.\,(\ref{eq:k1eff}) being achieved when $\sqrt{F_{j}}\propto u_{j}$, as indicated by the Cauchy--Schwarz inequality. For instance, when driving the system through $M$ input channels to the comb lines around
the center ($u_{j}\approx u_{p}$) simultaneously, with the driving amplitudes are comparable for all channels ($F_{j}\approx\frac{1}{M}F_{\mathrm{tot}}$),  we have $\kappa_{1}^{\mathrm{eff}}\approx M\kappa_{p,1}$ and achieves a $\sqrt{M}$-fold enhancement of soliton conversion efficiency. However, such an approach requires a pre-generated coherent comb, or a pulse
laser with the repetition rate matching that of the microsoliton, which imposes difficulties
in practice, especially for high-repetition-rate comb.

As the underlying mechanism of the efficiency limitations is revealed, we found that the generalized critical coupling condition of the collective state is very difficult to be achieved by the conventional couplers. Therefore, we proposed a new approach, i.e. the nonlinear couplers, to break these limitations, which hold more functionalities beyond the conventional linear couplers and are more flexible in experiments. The concept of the nonlinear coupler is schematically illustrated in Fig.~\ref{Fig1}(D), where the pump laser could excite the collective state indirectly through nonlinear frequency mixing processes, thus the conversion from the input CW laser to the soliton state could be power-dependent and also be parallel through multiple channels simultaneously.

\section{Nonlinear coupler: collective coupling}
\begin{figure}
	\begin{centering}
		\includegraphics[width=1\columnwidth]{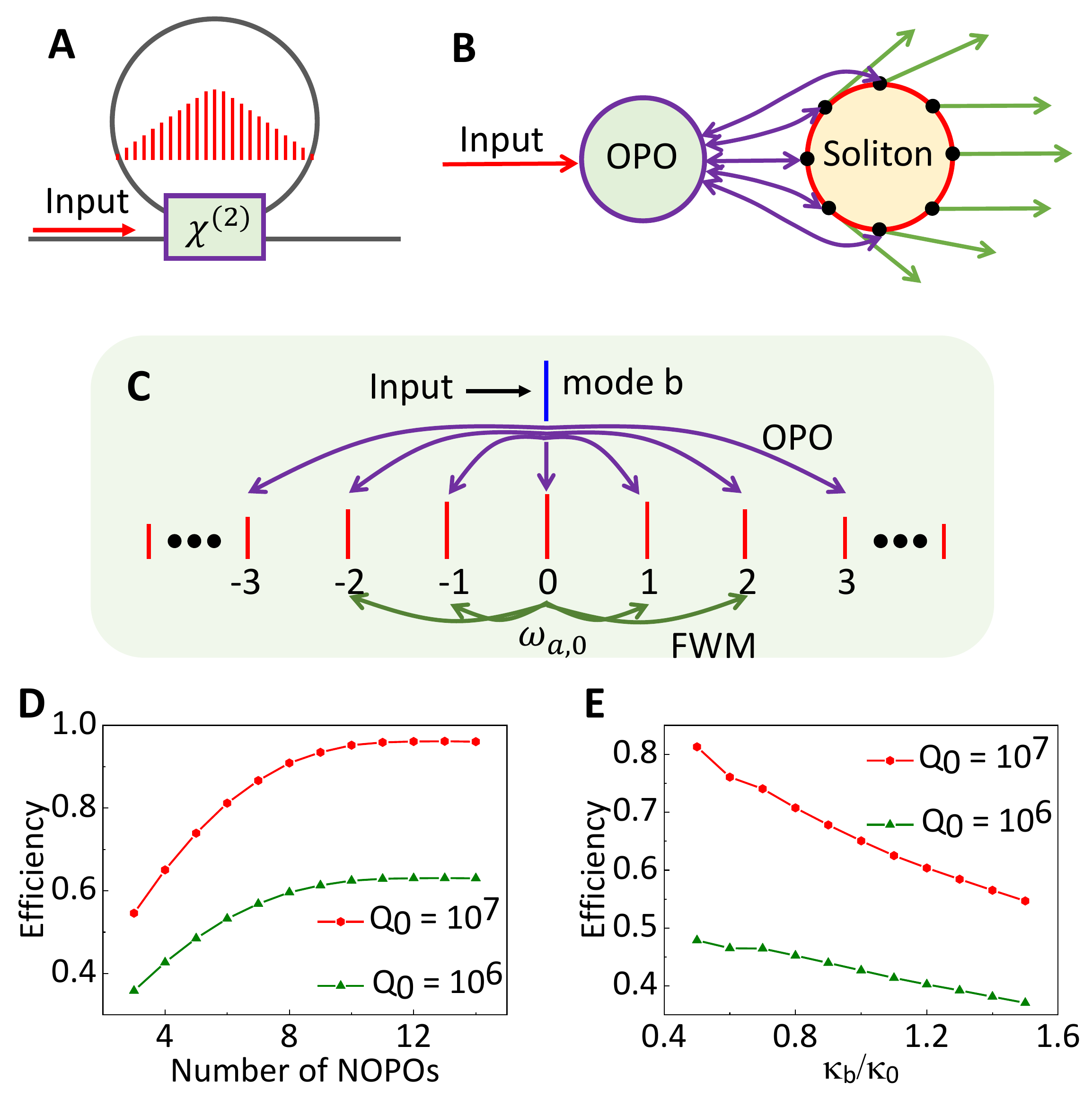}
		\par\end{centering}
	\caption{\textbf{Collective enhancement of soliton conversion efficiency by a nonlinear coupler}. (\textbf{A}) Schematic of a soliton cavity nonlinearly coupled to the input by $\chi^{(2)}$ material. (\textbf{B}) Input-output relation modified by OPO. The input field couples with multiple pairs of comb lines by $\chi^{(2)}$ interaction, as long as the phase-matching condition is fulfilled. (\textbf{C}) Illustration of the interactions between the input field and the cavity modes involved in the nonlinear coupler. The nonlinear coupler can be realized by introducing an intermediate mode (blue), which couples with different pairs of soliton modes (red). By eliminating the intermediate mode, the photon flux of the CW pump field is transferred to the soliton state through multiple channels via various
		OPOs. (\textbf{D})-(\textbf{E}) The dependence of the comb conversion efficiency
		on the number of mode-pairs that interact with $b$ by OPO (\textbf{D}) and
		the mode dissipation rate $\kappa_{b}$ (\textbf{E}). The collective enhancement
		is demonstrated by the positive relationship between the efficiency
		and number of NOPOs. Parameters: the nonlinear coupling strength $g_{2}=80\:$kHz, the normalization factor $\kappa_{0}=\frac{\omega_{b}}{2Q_{b}}$, and the loaded quality factors $Q_{a}=3\times10^{5}$, $Q_{b}=1\times10^{5}$.}
	
	\label{Fig2}
\end{figure}

As an example, we propose a nonlinear coupler to simultaneously excite various comb modes through the parametric driving using a CW laser, as shown by Fig.\,\ref{Fig2}(A). The waveguide couples with the cavity via $\chi^{(2)}$ interaction while the linear coupling between them is forbidden. Figure\,\ref{Fig2}(B) illustrates the modified input-output relation with collective
coupling between multiple comb modes and the input field. In practice, it is convenient to realize an efficient nonlinear coupler by introducing an ancillary cavity mode between the input field and the comb state for enhanced nonlinear coupling. In this example, the OPO is utilized to indirectly drive the soliton state, as the OPO process could naturally couple a CW laser to many pairs of comb line. To be more specific, as shown by Fig.\,\ref{Fig2}(C), the pump mode $b$ couples to the soliton state by the degenerate parametric interaction with
$a_{0}$ and also by the non-degenerate interaction (NOPO) with mode-pair $a_{\pm j}$ with $j\in\pm\{1\ldots m\}$. The corresponding interaction Hamiltonian reads $H_{\mathrm{OPO}}=\sum_{l}g_{2}\left(a_{j}a_{-j}b^{\dagger}+ba_{j}^{\dagger}a_{-j}^{\dagger}\right)$ for phase-matched $\chi^{(2)}$ nonlinear process.
Following a similar procedure of PFD (see the Supplementary Materials), the conversion
efficiency is derived as
\begin{eqnarray}
	\eta & = & \frac{\kappa_{a,1}}{\kappa_{a}}\frac{4g_{2}\sum_{j\in C}\mathrm{Im}[\alpha_{j}^{*}\alpha_{-j}^{*}\beta]}{P_{in}/\hbar\omega_{a,0}},\label{eq:dyOPO}
	\label{eq:OPOeff}
\end{eqnarray}
with $C=\pm\{0,1,2...\}$, $\alpha_{j}$ is the amplitude of mode
$a_{j}$, and $\beta=(-ig_{2}\alpha_{0}^{2}-ig_{2}\sum_{j\in C}\alpha_{-j}\alpha_{j}+\varepsilon_{b})/\kappa_{b}$, and the frequency of $a_0$ satisfies $\omega_{a,0}\approx\omega_{p}$ for the degenerate OPO interaction.

To verify such collective enhancement by OPO-based nonlinear coupler {[}Eq.\,(\ref{eq:OPOeff}){]},
we numerically investigated the relationship between the conversion
efficiency and the number of mode-pairs in $C$ using parameters of
a typical AlN microcavity~\cite{Bruch2020}. The numerical simulation
is performed using the mode expansion model\,\cite{PhysRevA.93.033820} (details for soliton simulation
in the Supplementary Materials). Figure\,\ref{Fig2}(D) shows the relationship between the
maximal conversion efficiency and number of mode-pairs $M$ involved in
the parametric interaction. As expected, the efficiency $\eta$ increases with
the mode number $M$, and eventually $\eta$ saturates for large $M$, because $\alpha_{j}$
decreases along with the wings of the comb and the contribution to $\eta$
of large $j$ can be neglected.

It should be noted that, even in absence of the collective effect that only one pair of mode couples with $b$, the nonlinear coupler could still provides an enhancement. For simplicity, we only consider the degenerate OPO process $C=\{0\}$, and the resulting conversion efficiency reduces to
\begin{eqnarray}
	\eta^{\{0\}} & = & \frac{\kappa_{a,1}}{\kappa_{a}}\frac{\kappa_{b,1}}{\kappa_{b}}\frac{8g_{2}\alpha^{2}}{\varepsilon_{b}^{2}}\left(\varepsilon_{b}-g_{2}\alpha^{2}\right).\label{eq:single}
\end{eqnarray}
Similar to the linear coupler cases, $\eta^{\{0\}}$ can also achieves
the optimal conversion efficiency $\frac{\kappa_{a,1}}{\kappa_{a}}\frac{\kappa_{b,1}}{\kappa_{b}}$
[Eq.\,(\ref{eq:maxefficiency-1})] when the critical coupling condition of Eq.\,($\ref{eq:ccm}$) is meet for $g_{2}=\frac{\varepsilon_{b}}{2\alpha^{2}}$.
Compared with Eq.\,(\ref{eq:EnergyDyn}) for the linear driving
case, the effective external coupling rate is magnified by a factor
of $\frac{4g_{2}\alpha}{\kappa_{b}}$. This is appealing in high-Q
(small $\kappa_{b}$) microresonator made by strong $\chi^{(2)}$
materials, such as AlN~\cite{Bruch:19}, LN~\cite{Lu:21,jankowski2020ultrabroadband}, and gallium arsenide~\cite{savanier2013near}, in which the OPO can increase the
conversion efficiency for $\frac{2g_{2}\alpha}{\kappa_{b}}>1$. The
magnified factor $\propto1/\kappa_{b}$ also holds in the
collective enhancement by multiple NOPOs, which is verified by numerical
simulation in Fig.\,\ref{Fig2}(E). As $\kappa_{b}$ decreases,
the conversion efficiency increases from $55\:$\% to $88\:$\% for
intrinsic quality factor $Q_{0}=10^{7}$, demonstrating the linear
modification of conversion efficiency by the OPO. The calculation is performed
for $C=\{0,1,2,3,4\}$, and the trend holds for other $C$.

\section{Nonlinear coupler: self-adaptive}
\begin{figure*}[t]
	\begin{centering}
		\includegraphics[width=1\textwidth]{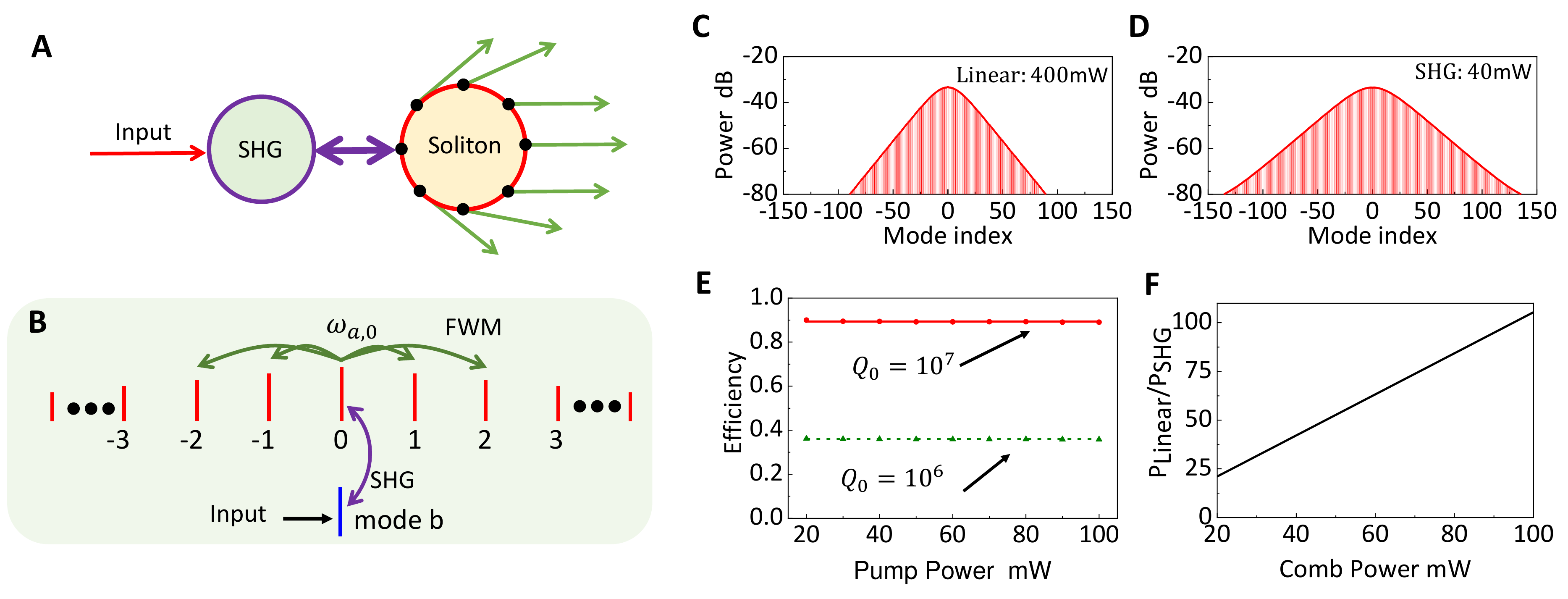}
		\par\end{centering}
	\caption{\textbf{Power-independent conversion efficiency by a self-adaptive nonlinear coupler}. (\textbf{A}) The generalized input-output relation modified by the second-harmonic generation (SHG). (\textbf{B}) Illustration of the interactions between the input field and the soliton state. An ancillary mode $b$ is used to mediate the SHG between the input field and a single comb line of the soliton state. 
		%is driven by the second harmonic generation between the intermediate mode of
		%frequency $\omega$ and the $a_{0}$ mode with frequency near $2\omega$.
		The SHG induces a pump-power-dependent external coupling rate between
		the waveguide and the soliton state. (\textbf{C})-(\textbf{D}) Simulated intracavity soliton spectral for linear coupler (400~mW) and  SHG-based nonlinear coupler (40~mW) under the same parameter of the soliton cavity. Even though the pump power in
		the SHG-based nonlinear coupler case is 10 times lower, the comb bandwidth is much
		larger, so is the conversion efficiency. The numerical simulations are performed based on the parameters of a typical AlN microring~\cite{Bruch2020}. (\textbf{E}) Power-independent relationship between the conversion efficiency and the pump power for the SHG nonlinear coupler. (\textbf{F}) The power-saving factor $P_{\mathrm{Linear}}/P_{\mathrm{SHG}}$, as the ratio of the required pump power of  SHG-based nonlinear coupler scheme to that of the linear coupler scheme for a target output comb power.}
	\label{Fig3}
\end{figure*}

\noindent The above analyses have resolved the
first efficiency limitation on the comb conversion by effectively amplifying the external coupling rate and keeping the total dissipation rate of the soliton state unchanged. However, for both the linear- and OPO-based nonlinear-coupler schemes, the effective external coupling rate $\kappa^{\mathrm{eff}}$ decreases with the bandwidth of the soliton state, and the efficiency is power-dependent [Fig.\,\ref{Fig1}(C)]. 
According to the unique nature of soliton state, an ideal coupler requires $\kappa_{p,1}$ to be adaptively variable with input power, i.e. be proportional to $P_{\mathrm{in}}$ to ensure a linear relationship between $S_{\mathrm{in}}$ and $P_{\mathrm{in}}$ or a $P_{\mathrm{in}}$-independent $\kappa_{1}^{\mathrm{eff}}$. By designing
an intermediate passive nonlinear coupler to modify the function $S_{\mathrm{in}}\left(P_{\mathrm{in}}\right)$,
this coupler could be realized self-adaptively, without requiring a real-time feedback control. 

Here, we provide an example of the nonlinear
coupler via SHG, which meets the self-adaptive
requirement due to the quadratic relationship between the input and
the second-harmonic light. The illustration of the modified input-output relation and configuration of the nonlinear interactions are depicted in Figs.\,\ref{Fig3}(A)-(B).
The nonlinear interaction between the comb mode $a_{0}$ and ancillary $b$ is described by the Hamiltonian of $H_{\mathrm{SHG}}=g_{2}\left(a_{0}b^{\dagger2}+b^{2}a_{0}^{\dagger}\right)$,
with mode $b$ driven by the input field following the driving term
$H_{d}=i\varepsilon_{b}\left(b^{\dagger}-b\right)$. According to the
PFD (see the Supplementary Materials for more details), the conversion efficiency is derived as
\begin{eqnarray}
	\eta & = & \frac{\kappa_{a,1}}{\kappa_{a}}\frac{8\kappa_{b,1}g_{2}\alpha}{\left(\kappa_{b}+2g_{2}\alpha\right)^{2}},\label{eq:effSHG}
\end{eqnarray}
which is independent of the pump power. It should be noted that, under the critical coupling condtion of Eq.\,($\ref{eq:ccm}$), the conversion efficiency $\eta$ also has a maximum of $\eta_{max}=\frac{\kappa_{a,1}}{\kappa_{a}}\frac{\kappa_{b,1}}{\kappa_{b}}$
for $\kappa_{b}=2g_{2}\alpha$, which is consistent with the analysis in Eq.\,(\ref{eq:maxefficiency-1}).

The analytical prediction of the SHG-based coupler is further validated
by numerical simulation for a practical AlN microresonator\,\cite{Bruch2020}.
Figures\,\ref{Fig3}(C)-(D) show the optical spectrum of a single-soliton
state for the conventional linear coupler scheme and the SHG-based nonlinear coupler schemes. Compared with Kerr soliton comb generated by the conventional linear coupler scheme [Fig.\,\ref{Fig3}(C)], it still maintains a $\mathrm{sech}^{2}$
profile, but has a much wider spectrum even with $10$ times lower
pump power. Since the $\eta$ is eventually limited by $\frac{\kappa_{b,1}}{\kappa_{b}}$
and $\frac{\kappa_{a,1}}{\kappa_{a}}$, which denote the excitation
and extraction efficiencies, respectively, the over-coupling condition
raises strict requirement on high intrinsic quality factor $Q_{0}$
for nonlinear frequency conversion towards near-unity efficiency. Then
we calculate the conversion efficiency for loaded cavity quality factors
$Q_{a}=6\times10^{5}$ and $Q_{b}=6\times10^{4}$, and $\chi^{(2)}$
nonlinear coupling rate $g_{2}/2\pi=140\:$kHz. As shown in Fig.\,\ref{Fig3}(E),
the conversion efficiency obtained by numerical simulation (dots)
is pump power-independent and maintains a high value near $90\:\%$
for $Q_{0}=10^{7}$ and $36\:\%$ for $Q_{0}=10^{6}$, both in
excellent agreement with the analytical result of Eq.\,(\ref{eq:effSHG}),
as shown by the solid lines. It is anticipated that the conversion
efficiency can be further promoted to approaching unity by increasing
the external coupling rates $\kappa_{a(b),1}$, albeit higher $\kappa_{a(b)}$, in turn, demands a higher $g_{2}$. To provide a fair comparison between
the adaptive nonlinear coupler and the conventional linear coupler,
we studied the required pump powers to generate a soliton with the
same output power. With all the parameters of the Kerr cavity fixed,
the power saving factor, defined as the ratio of the required input powers $F=P_{\mathrm{Linear}}/P_{\mathrm{SHG}}$
is plotted in Fig.\,\ref{Fig3}(F). The results show a reduction
of required input power by more than two orders for a high output comb power.

Besides the SHG, the mechanism of the self-adaptive coupler applies to any nonlinear process that builds a linear relationship between the input photon flux $S_{\mathrm{in}}$ and the input power $P_{\mathrm{in}}$. Interestingly, such a relation could be obtained by other nonlienar processes, such as the Raman scattering and Brillouin scattering, which exist in almost all kinds of dielectric materials. The self-adaptive coupler is realized by nonlinearly pumping the soliton state using the first-order Raman lasing of the ancillary mode. Following the same procedure of PFD, the pump-power-independent conversion efficiency is derived as (see the Supplementary Materials)
\begin{eqnarray}
	\eta & = & \frac{\kappa_{a,1}}{\kappa_{a}}\frac{4\kappa_{b,1}}{\left(\frac{\kappa_{b}}{\sqrt{g_{R}^{2}\alpha^2/\kappa_R}}+\sqrt{\frac{g_{R}^{2}\alpha^2}{\kappa_{R}}}\right)^{2}},\label{eq:effRaman}
\end{eqnarray}
where $g_{R}$ and $\kappa_{R}$ are the coupling strength and gain bandwidth of Raman scattering, respectively. Under the condition of Eq.\,($\ref{eq:ccm}$), the conversion efficiency also achieves the maximum $\frac{\kappa_{a,1}}{\kappa_{a}}\frac{\kappa_{b,1}}{\kappa_{b}}$ when the dissipation rate of the ancillary mode is set to a pump-power-independent value as $\kappa_{b}=\frac{g_{R}^{2}\alpha}{\kappa_{R}}$. 

\section{Discussion}

The proposed nonlinear coupler to break the efficiency limitations of micro-soliton generation is experimentally feasible for a variety of integrated platforms made by different materials. Firstly, the efficient SHG and OPO have been experimentally demonstrated in integrated AlN and LN microresonators~\cite{Bruch:19,lu2020toward,Lu:21}, as well as their roles in generating frequency combs\,\cite{PhysRevLett.124.203902,amiune2021optical,englebert2021parametrically}.
Here, a conversion efficiency over $90\%$ is predicted for AlN {[}Fig.\,\ref{Fig3}(D){]}, and it could be even higher for LN because of higher $\chi^{(2)}$ nonlinearity and quality factor~\cite{lu2020toward,PhysRevApplied.16.064004,zhang2019fabrication}. Secondly, an effective $\chi^{(2)}$ nonlinearity can be constructed from $\chi^{(3)}$ process by an additional DC bias field, as demonstrated in SiN microresonator~\cite{Lu2021} and silicon waveguide~\cite{Timurdogan2017}.
Thirdly, the optimal self-adaptive nonlinear coupler could also be realized by other nonlinear processes such as the Raman scattering [Eq.~(\ref{eq:effRaman})] and Brillouin scattering [see Supplementary Materials], which could be found in almost all dielectric materials without stringent design for phase-matching. For example, it has recently been demonstrated that the soliton comb is achievable by Raman lasing$\:$\cite{Yang2017} and Brillouin lasing~\cite{bai2020brillouin}, we would expect an observation of high power conversion efficiency based on these works. Therefore, the soliton generation with near-unity efficiency that be insensitive to the pump power could be realized in almost all material platforms.

In conclusion, the underlying physics that imposes the efficiency limitations of soliton frequency comb generation is studied under a unified theoretical framework, in which the general critical coupling for multimode nonlinear optics system is derived. It is revealed that the key for achieving high conversion efficiency is to compensate the weighted dissipation rate of a soliton state, which requires a power-dependent strong external coupling to a selected pump mode. To break the limitations, we propose nonlinear couplers to drive the soliton indirectly with continuous-wave lasers. In particular, the pump-power-independent self-adaptive coupler can be realized via second-harmonic generation or Raman effects, which is feasible for experiments and promises a high-efficiency and robust soliton state generation on various platforms. The approach developed in this work is also applicable to other frequency comb states (e.g. multiple solitons, primary comb state) and other nonlinear frequency conversion processes~\cite{rueda2019resonant} that involve a steady-state of multiple modes, and therefore we believed it worth further theoretical and experimental insights.

\smallskip{}

\textbf{Acknowledgments}\\We thank Dr. Alexander Bruch for discussions. This work was funded by the National Natural Science Foundation of China (Grants Nos.~11934012, 11874342, 11904316, 11922411, 61690192, U21A20433, 12104441, U21A6006), Anhui Provincial Natural Science Foundation (Grant No.\,2008085QA34, 2108085MA22), and Major Scientific Project of Zhejiang Laboratory (No.\,2020LC0AD01). ML and CLZ were also supported by the Fundamental Research Funds for the Central Universities, and the State Key Laboratory of Advanced Optical Communication Systems and Networks. The numerical calculations in this paper have been done on the supercomputing system in the Supercomputing Center of University of Science and Technology of China.

%\bibliographystyle{aps}
%\bibliography{EfficientComb}

\begin{thebibliography}{10}
	\providecommand{\url}[1]{\texttt{#1}}
	\providecommand{\urlprefix}{URL }
	\providecommand{\eprint}[2][]{\url{#2}}
	
	\bibitem{fortier201920}
	T.~Fortier and E.~Baumann, 20 years of developments in optical frequency comb
	technology and applications, Communications Physics \textbf{2}, 1 (2019).
	
	\bibitem{kippenberg2018dissipative}
	T.~J. Kippenberg, A.~L. Gaeta, M.~Lipson, and M.~L. Gorodetsky, Dissipative
	kerr solitons in optical microresonators, Science \textbf{361}, eaan8083
	(2018).
	
	\bibitem{PhysRevA.89.063814}
	C.~Godey, I.~V. Balakireva, A.~Coillet, and Y.~K. Chembo, Stability analysis of
	the spatiotemporal lugiato-lefever model for kerr optical frequency combs in
	the anomalous and normal dispersion regimes, Phys. Rev. A \textbf{89}, 063814
	(2014).
	
	\bibitem{Cole2017}
	D.~C. Cole, E.~S. Lamb, P.~Del?Haye, S.~A. Diddams, and S.~B. Papp, Soliton
	crystals in kerr resonators, Nature Photonics \textbf{11}, 671 (2017).
	
	\bibitem{karpov2019dynamics}
	M.~Karpov, M.~H. Pfeiffer, H.~Guo, W.~Weng, J.~Liu, and T.~J. Kippenberg,
	Dynamics of soliton crystals in optical microresonators, Nature Physics
	\textbf{15}, 1071 (2019).
	
	\bibitem{Lu2021a}
	Z.~Lu, H.-J. Chen, W.~Wang, L.~Yao, Y.~Wang, Y.~Yu, B.~E. Little, S.~T. Chu,
	Q.~Gong, W.~Zhao, X.~Yi, Y.-F. Xiao, and W.~Zhang, Synthesized soliton
	crystals, Nature Communications \textbf{12}, 3179 (2021).
	
	\bibitem{moss2013new}
	D.~J. Moss, R.~Morandotti, A.~L. Gaeta, and M.~Lipson, New cmos-compatible
	platforms based on silicon nitride and hydex for nonlinear optics, Nature
	photonics \textbf{7}, 597 (2013).
	
	\bibitem{PhysRevLett.114.053901}
	S.-W. Huang, H.~Zhou, J.~Yang, J.~F. McMillan, A.~Matsko, M.~Yu, D.-L. Kwong,
	L.~Maleki, and C.~W. Wong, Mode-locked ultrashort pulse generation from
	on-chip normal dispersion microresonators, Phys. Rev. Lett. \textbf{114},
	053901 (2015).
	
	\bibitem{herr2014temporal}
	T.~Herr, V.~Brasch, J.~D. Jost, C.~Y. Wang, N.~M. Kondratiev, M.~L. Gorodetsky,
	and T.~J. Kippenberg, Temporal solitons in optical microresonators, Nature
	Photonics \textbf{8}, 145 (2014).
	
	\bibitem{gong2018high}
	Z.~Gong, A.~Bruch, M.~Shen, X.~Guo, H.~Jung, L.~Fan, X.~Liu, L.~Zhang, J.~Wang,
	J.~Li \emph{et~al.}, High-fidelity cavity soliton generation in crystalline
	aln micro-ring resonators, Optics letters \textbf{43}, 4366 (2018).
	
	\bibitem{yi2015soliton}
	X.~Yi, Q.-F. Yang, K.~Y. Yang, M.-G. Suh, and K.~Vahala, Soliton frequency comb
	at microwave rates in a high-q silica microresonator, Optica \textbf{2}, 1078
	(2015).
	
	\bibitem{gong2020photonic}
	Z.~Gong, M.~Li, X.~Liu, Y.~Xu, J.~Lu, A.~Bruch, J.~B. Surya, C.~Zou, and H.~X.
	Tang, Photonic dissipation control for kerr soliton generation in strongly
	raman-active media, Physical Review Letters \textbf{125}, 183901 (2020).
	
	\bibitem{gong2020near}
	Z.~Gong, X.~Liu, Y.~Xu, and H.~X. Tang, Near-octave lithium niobate soliton
	microcomb, Optica \textbf{7}, 1275 (2020).
	
	\bibitem{he2020perfect}
	Y.~He, J.~Ling, M.~Li, and Q.~Lin, Perfect soliton crystals on demand, Laser \&
	Photonics Reviews \textbf{14}, 1900339 (2020).
	
	\bibitem{guo2017universal}
	H.~Guo, M.~Karpov, E.~Lucas, A.~Kordts, M.~H. Pfeiffer, V.~Brasch, G.~Lihachev,
	V.~E. Lobanov, M.~L. Gorodetsky, and T.~J. Kippenberg, Universal dynamics and
	deterministic switching of dissipative kerr solitons in optical
	microresonators, Nature Physics \textbf{13}, 94 (2017).
	
	\bibitem{xue2015mode}
	X.~Xue, Y.~Xuan, Y.~Liu, P.-H. Wang, S.~Chen, J.~Wang, D.~E. Leaird, M.~Qi, and
	A.~M. Weiner, Mode-locked dark pulse kerr combs in normal-dispersion
	microresonators, Nature Photonics \textbf{9}, 594 (2015).
	
	\bibitem{pasquazi2018micro}
	A.~Pasquazi, M.~Peccianti, L.~Razzari, D.~J. Moss, S.~Coen, M.~Erkintalo, Y.~K.
	Chembo, T.~Hansson, S.~Wabnitz, P.~Del\^{a}Haye \emph{et~al.}, Micro-combs: A
	novel generation of optical sources, Physics Reports \textbf{729}, 1 (2018).
	
	\bibitem{griffith2015silicon}
	A.~G. Griffith, R.~K. Lau, J.~Cardenas, Y.~Okawachi, A.~Mohanty, R.~Fain,
	Y.~H.~D. Lee, M.~Yu, C.~T. Phare, C.~B. Poitras \emph{et~al.}, Silicon-chip
	mid-infrared frequency comb generation, Nature communications \textbf{6},
	6299 (2015).
	
	\bibitem{foltynowicz2011optical}
	A.~Foltynowicz, P.~Mas{\l}owski, T.~Ban, F.~Adler, K.~Cossel, T.~Briles, and
	J.~Ye, Optical frequency comb spectroscopy, Faraday discussions \textbf{150},
	23 (2011).
	
	\bibitem{picque2019frequency}
	N.~Picqu{\'e} and T.~W. H{\"a}nsch, Frequency comb spectroscopy, Nature
	Photonics \textbf{13}, 146 (2019).
	
	\bibitem{marin2017microresonator}
	P.~Marin-Palomo, J.~N. Kemal, M.~Karpov, A.~Kordts, J.~Pfeifle, M.~H. Pfeiffer,
	P.~Trocha, S.~Wolf, V.~Brasch, M.~H. Anderson \emph{et~al.},
	Microresonator-based solitons for massively parallel coherent optical
	communications, Nature \textbf{546}, 274 (2017).
	
	\bibitem{wang2020quantum}
	F.-X. Wang, W.~Wang, R.~Niu, X.~Wang, C.-L. Zou, C.-H. Dong, B.~E. Little,
	S.~T. Chu, H.~Liu, P.~Hao \emph{et~al.}, Quantum key distribution with
	on-chip dissipative kerr soliton, Laser \& Photonics Reviews \textbf{14},
	1900190 (2020).
	
	\bibitem{obrzud2019microphotonic}
	E.~Obrzud, M.~Rainer, A.~Harutyunyan, M.~H. Anderson, J.~Liu, M.~Geiselmann,
	B.~Chazelas, S.~Kundermann, S.~Lecomte, M.~Cecconi \emph{et~al.}, A
	microphotonic astrocomb, Nature Photonics \textbf{13}, 31 (2019).
	
	\bibitem{ghelfi2014fully}
	P.~Ghelfi, F.~Laghezza, F.~Scotti, G.~Serafino, A.~Capria, S.~Pinna, D.~Onori,
	C.~Porzi, M.~Scaffardi, A.~Malacarne \emph{et~al.}, A fully photonics-based
	coherent radar system, Nature \textbf{507}, 341 (2014).
	
	\bibitem{liu2020photonic}
	J.~Liu, E.~Lucas, A.~S. Raja, J.~He, J.~Riemensberger, R.~N. Wang, M.~Karpov,
	H.~Guo, R.~Bouchand, and T.~J. Kippenberg, Photonic microwave generation in
	the x-and k-band using integrated soliton microcombs, Nature Photonics
	\textbf{14}, 486 (2020).
	
	\bibitem{spencer2018optical}
	D.~T. Spencer, T.~Drake, T.~C. Briles, J.~Stone, L.~C. Sinclair, C.~Fredrick,
	Q.~Li, D.~Westly, B.~R. Ilic, A.~Bluestone \emph{et~al.}, An
	optical-frequency synthesizer using integrated photonics, Nature
	\textbf{557}, 81 (2018).
	
	\bibitem{Papp:14}
	S.~B. Papp, K.~Beha, P.~Del'Haye, F.~Quinlan, H.~Lee, K.~J. Vahala, and S.~A.
	Diddams, Microresonator frequency comb optical clock, Optica \textbf{1}, 10
	(2014).
	
	\bibitem{Feldmann2021}
	J.~Feldmann, N.~Youngblood, M.~Karpov, H.~Gehring, X.~Li, M.~Stappers,
	M.~Le~Gallo, X.~Fu, A.~Lukashchuk, A.~S. Raja, J.~Liu, C.~D. Wright,
	A.~Sebastian, T.~J. Kippenberg, W.~H.~P. Pernice, and H.~Bhaskaran, Parallel
	convolutional processing using an integrated photonic tensor core, Nature
	\textbf{589}, 52 (2021).
	
	\bibitem{bao2014nonlinear}
	C.~Bao, L.~Zhang, A.~Matsko, Y.~Yan, Z.~Zhao, G.~Xie, A.~M. Agarwal, L.~C.
	Kimerling, J.~Michel, L.~Maleki \emph{et~al.}, Nonlinear conversion
	efficiency in kerr frequency comb generation, Optics Letters \textbf{39},
	6126 (2014).
	
	\bibitem{jang2020universal}
	J.~K. Jang, Y.~Okawachi, X.~Ji, C.~Joshi, M.~Lipson, and A.~L. Gaeta, Universal
	conversion efficiency scaling with free-spectral-range for soliton kerr
	combs, in \emph{2020 Conference on Lasers and Electro-Optics (CLEO)}  (IEEE
	2020), pp. 1--2.
	
	\bibitem{xue2019super}
	X.~Xue, X.~Zheng, and B.~Zhou, Super-efficient temporal solitons in mutually
	coupled optical cavities, Nature Photonics \textbf{13}, 616 (2019).
	
	\bibitem{Bruch2020}
	A.~W. Bruch, X.~Liu, Z.~Gong, J.~B. Surya, M.~Li, C.-L. Zou, and H.~X. Tang,
	Pockels soliton microcomb, Nature Photonics  (2020).
	
	\bibitem{obrzud2017temporal}
	E.~Obrzud, S.~Lecomte, and T.~Herr, Temporal solitons in microresonators driven
	by optical pulses, Nature Photonics \textbf{11}, 600 (2017).
	
	\bibitem{malinowski2017optical}
	M.~Malinowski, A.~Rao, P.~Delfyett, and S.~Fathpour, Optical frequency comb
	generation by pulsed pumping, APL Photonics \textbf{2}, 066101 (2017).
	
	\bibitem{anderson2018achieving}
	M.~Anderson, R.~Bouchand, E.~Obrzud, J.~Liu, S.~Karlen, E.~Lucas, S.~Lecomte,
	T.~Herr, and T.~J. Kippenberg, Achieving efficient conversion and broadband
	operation in pulse-driven kerr microresonators, in \emph{Frontiers in Optics}
	(Optical Society of America 2018), pp. FW7B--4.
	
	\bibitem{guo2018efficient}
	X.~Guo, C.-L. Zou, H.~Jung, Z.~Gong, A.~Bruch, L.~Jiang, and H.~X. Tang,
	Efficient generation of a near-visible frequency comb via cherenkov-like
	radiation from a kerr microcomb, Physical Review Applied \textbf{10}, 014012
	(2018).
	
	\bibitem{walls2007quantum}
	D.~F. Walls and G.~J. Milburn, \emph{Quantum optics}  (Springer Science \&
	Business Media 2007).
	
	\bibitem{PhysRevA.93.033820}
	Y.~K. Chembo, Quantum dynamics of kerr optical frequency combs below and above
	threshold: Spontaneous four-wave mixing, entanglement, and squeezed states of
	light, Phys. Rev. A \textbf{93}, 033820 (2016).
	
	\bibitem{Bruch:19}
	A.~W. Bruch, X.~Liu, J.~B. Surya, C.-L. Zou, and H.~X. Tang, On-chip
	$\chi^{(2)}$ microring optical parametric oscillator, Optica \textbf{6}, 1361
	(2019).
	
	\bibitem{Lu:21}
	J.~Lu, A.~A. Sayem, Z.~Gong, J.~B. Surya, C.-L. Zou, and H.~X. Tang,
	Ultralow-threshold thin-film lithium niobate optical parametric oscillator,
	Optica \textbf{8}, 539 (2021).
	
	\bibitem{jankowski2020ultrabroadband}
	M.~Jankowski, C.~Langrock, B.~Desiatov, A.~Marandi, C.~Wang, M.~Zhang, C.~R.
	Phillips, M.~Lon{\v{c}}ar, and M.~Fejer, Ultrabroadband nonlinear optics in
	nanophotonic periodically poled lithium niobate waveguides, Optica
	\textbf{7}, 40 (2020).
	
	\bibitem{savanier2013near}
	M.~Savanier, C.~Ozanam, L.~Lanco, X.~Lafosse, A.~Andronico, I.~Favero,
	S.~Ducci, and G.~Leo, Near-infrared optical parametric oscillator in a iii-v
	semiconductor waveguide, Applied Physics Letters \textbf{103}, 261105 (2013).
	
	\bibitem{lu2020toward}
	J.~Lu, M.~Li, C.-L. Zou, A.~Al~Sayem, and H.~X. Tang, Toward 1\% single-photon
	anharmonicity with periodically poled lithium niobate microring resonators,
	Optica \textbf{7}, 1654 (2020).
	
	\bibitem{PhysRevLett.124.203902}
	J.~Szabados, D.~N. Puzyrev, Y.~Minet, L.~Reis, K.~Buse, A.~Villois, D.~V.
	Skryabin, and I.~Breunig, Frequency comb generation via cascaded second-order
	nonlinearities in microresonators, Phys. Rev. Lett. \textbf{124}, 203902
	(2020).
	
	\bibitem{amiune2021optical}
	N.~Amiune, D.~N. Puzyrev, V.~V. Pankratov, D.~V. Skryabin, K.~Buse, and
	I.~Breunig, Optical-parametric-oscillation-based $\chi$ (2) frequency comb in
	a lithium niobate microresonator, Optics Express \textbf{29}, 41378 (2021).
	
	\bibitem{englebert2021parametrically}
	N.~Englebert, F.~De~Lucia, P.~Parra-Rivas, C.~M. Arab{\'\i}, P.-J. Sazio, S.-P.
	Gorza, and F.~Leo, Parametrically driven kerr cavity solitons, Nature
	Photonics \textbf{15}, 857 (2021).
	
	\bibitem{PhysRevApplied.16.064004}
	J.-Y. Chen, Z.~Li, Z.~Ma, C.~Tang, H.~Fan, Y.~M. Sua, and Y.-P. Huang, Photon
	conversion and interaction in a quasi-phase-matched microresonator, Phys.
	Rev. Applied \textbf{16}, 064004 (2021).
	
	\bibitem{zhang2019fabrication}
	J.~Zhang, Z.~Fang, J.~Lin, J.~Zhou, M.~Wang, R.~Wu, R.~Gao, and Y.~Cheng,
	Fabrication of crystalline microresonators of high quality factors with a
	controllable wedge angle on lithium niobate on insulator, Nanomaterials
	\textbf{9}, 1218 (2019).
	
	\bibitem{Lu2021}
	X.~Lu, G.~Moille, A.~Rao, D.~A. Westly, and K.~Srinivasan, Efficient
	photoinduced second-harmonic generation in silicon nitride photonics, Nature
	Photonics \textbf{15}, 131 (2021).
	
	\bibitem{Timurdogan2017}
	E.~Timurdogan, C.~V. Poulton, M.~J. Byrd, and M.~R. Watts, Electric
	field-induced second-order nonlinear optical effects in silicon waveguides,
	Nature Photonics \textbf{11}, 200 (2017).
	
	\bibitem{Yang2017}
	Q.-F. Yang, X.~Yi, K.~Y. Yang, and K.~Vahala, Stokes solitons in optical
	microcavities, Nature Physics \textbf{13}, 53 (2017).
	
	\bibitem{bai2020brillouin}
	Y.~Bai, M.~Zhang, Q.~Shi, S.~Ding, Y.~Qin, Z.~Xie, X.~Jiang, and M.~Xiao,
	Brillouin-kerr soliton frequency combs in an optical microresonator, Physical
	Review Letters \textbf{126}, 063901 (2021).
	
	\bibitem{rueda2019resonant}
	A.~Rueda, F.~Sedlmeir, M.~Kumari, G.~Leuchs, and H.~G. Schwefel, Resonant
	electro-optic frequency comb, Nature \textbf{568}, 378 (2019).
	
\end{thebibliography}

\end{document}